
\magnification=\magstep1 \hsize=16 true cm
\vsize=24 true cm
\quad\quad\quad\quad\quad\quad\quad\quad \quad\quad
\quad\quad\quad\quad\quad\quad\quad\quad\quad\quad\quad
\quad\quad\quad\quad\quad\ BIR-QMW/PH/91-1.
\vskip 1.5cm
\centerline{\bf {Hidden Kac-Moody symmetry}}
\centerline{\bf{and 2D Quantum Supergravity}}
\vskip 4cm
\centerline {\it{W.A. Sabra}}
\centerline{\it{Physics Department}}
\centerline{\it{Birkbeck College}}
\centerline{\it{University of London}}
\centerline{\it{Malet Street}}
\centerline{\it{ London WC1E 7HX}}
\vskip 2mm
\centerline{\it and}
\vskip 2mm
\centerline{\it{Physics Department}}
\centerline{\it{Queen Mary and Westfield College}}
\centerline{\it{University of London}}
\centerline{\it{Mile End Road}}
\centerline{\it{London E1 4NS}}
\centerline{\it{United Kingdom}}
\vskip 4cm

\centerline{ABSTRACT}
We examine the relation between Polyakov's formulation of two
dimensional supergravity and gauged Wess-Zumino-Novikov-Witten
models.

\vfill\eject

\centerline{ \bf{1. Introduction}}
\vskip 0.2 cm
The subject of two dimensional (super)conformal field theory
has connections with many active areas of both theoretical physics
and pure mathematics. In particular, two dimensional (super)conformal
field  theory coupled to two dimensional (super)gravity is not only
of interest  as a tractable theory of quantum (super)gravity but
an essential ingredient in the formulation of off-critical
(super)string theories in the description of elementary  particles.
{}From the point of view of statistical mechanics, two  dimensional
quantum gravity describes the critical theory at a second order phase
transition point on a random lattice.

The subject was transformed by Polyakov's treatment of two
dimensional gravity as an induced theory arising from quantum
averaging over some underlying matter degrees of freedom coupling to
a gravitational field via their stress-energy tensor
[1].

It is clear from the work of Polyakov $et$ $al.$ [1], that one
can gain a more understanding of the problem of quantizing
two dimensional quantum gravity  if,
instead of the conformal gauge, one chooses alternatively a \lq\lq
chiral" gauge. For this choice, there are analogies one can make
between two dimensional quantum gravity
and the Wess-Zumino-Novikov-Witten (WZNW) model [16]. This then led
Polyakov to the remarkable discovery that the theory  possesses an
associated chiral current algebra based upon the non-compact group
$SL(2,R)$.  This result was obtained by an explicit calculation of
the correlation functions for the gravitational field. A
generalization of this result to both the $N=1$ and $N=2$
supersymmetric case is possible [1,7], yielding  graded versions of
$SL(2,R)$, the orthogonal symplectic groups $OSp(1,2)$ and
$OSp(2,2)$ respectively.

Recently, there has been a lot of progress in understanding
the symmetries of Polyakov's two dimensional gravity and
supergravities in terms of the coadjoint orbits  method (for a
review see [11]) and the gauged Wess-Zumino-Novikov-Witten models.
Alekseev and Shatashvili [2] constructed a geometric action
on a particular coadjoint orbit of the Virasoro group
[3]. The action they derived is,
$$S_{vir}(F)={k\over 2\pi}\int d2z{\partial_{\bar z}
F\over{(\partial_z  F)}3}\Big(\partial_z F\partial_z3
F-2{(\partial_z2 F)}2\Big),\eqno(1.1)$$
where $\partial_z= {\partial/\partial z}$, $\partial_{\bar z}=
\partial/\partial_{\bar z}.$ \footnote *{we parametrize the two
dimensional space time with coordinates $(x, t)$ by
$z=t+x$ and $\bar z=t-x$.}  The action $S_{vir}(F)$ is
invariant under the $SL(2,R)$ transformation symmetry,  $$F(z,\bar
z)\rightarrow {d(\bar z)F+b(\bar z)\over c(\bar z)F+a(\bar z)};\qquad
ad-bc=1.\eqno(1.2)$$ In addition, in reference [2] it was shown that
the action (1.1) is classically equivalent to a Borel gauged WZNW
model with the gauge group $SL(2,R)$. The equivalence at the quantum
level was demonstrated by Bershadsky and Ooguri [4] by showing that
the representations of the conformal field theory are those of the
constrained $SL(2,R)$ WZNW model.

The two dimensional quantum gravity action can be
obtained from the geometric action (1.1) through the following
transformation, $$F(f(z,\bar z), \bar z)=z,\eqno(1.3)$$  and is
expressed in terms of the scalar field $f$ as,
$$S_{{grav}}(h_{\bar z\bar z})=-{k\over 2\pi}\int d2z
\Big({\partial_{z}2f\partial_z\partial_{\bar z}
f\over(\partial_{ z} f)2}-
{{(\partial_{z}2f)}2\partial_{\bar z} f\over
{(\partial_{z} f)}3}\Big), \quad h_{\bar z\bar
z}={\partial_{\bar z}f\over\partial_{z}f},\eqno(1.4)$$ where
$h_{{\bar z}{\bar z}}$ is the component of the gravitational field
surviving the light-cone gauge.

With regard to the supersymmetric extensions of these results, an
 $(1,0)$ supersymmetric geometric action has been formulated in
[5,6] on the coadjoint orbit (of purely central extension) of the
$N=1$ superVirasoro group.  The authors also derive an
action  for (1,0) supergravity by gauging the Borel subgroup of the
$(0,1)$ supersymmetric $SL(2,R)$ WZNW model. However,
the supergravity  action they obtained does not have the $OSp(1,2)$
current algebra but instead an $N=1$ supersymmetric extension of the
$SL(2,R)$  current algebra.
 This formulation of $(1,0)$ supergravity
[5,6], simply corresponds to a different gauge choice in the
$(1,0)$ supergravity theory, and is explained as follows.
In $(1,0)$ supergravity  the covariant derivatives are
expressed as,
$$\nabla_A=E_AMD_M+\omega_A M,\eqno(1.5)$$
where $E_AM$ are the vielbiens and $\omega_A$ is the connection.
The torsion and curvature are determined by the graded commutators
of the  covariant derivatives. In order to reduce the component
fields of supergravity, one must introduce a set of algebraic
constraints on these superfields. For $(1,0)$
supergravity, the unconstrained fields [23] are, in
component form, $$\eqalign{H{\bar
z}_{\theta}&=\rho_{\theta}{\bar z}+\theta h{\bar z}_{z},\cr
H{z}_{\bar z}
&=h{z}_{\bar z}+\theta\psi{\theta}_{\bar z},\cr
S&=h+\theta\psi{\theta}_{\bar z}.\cr}\eqno(1.6)$$

The component fields $\{h{z}_{\bar z}, h{\bar z}_{z}, h\}$
represent  the graviton, $\{\psi{\theta}_{\bar z},
\psi{\theta}_{z}\}$ represent the  gravitino and $\rho{\bar
z}_{\theta}$ is a pure gauge degree of freedom. Choosing the chiral
gauge, $H{\bar z}_\theta=0$, $S=1$, gives an action  whose degrees
of freedom are the  graviton and the gravitino, and the equation of
motion derived from  the anomaly equation [7,25] is,
$$\partial2_{z}D_{\theta}H{z}_{\bar z}=0.\eqno(1.7)$$
The symmetry of the theory in the above chiral gauge was
derived in [7] where it was shown that the system
possesses an $OSp(1,2)$ current algebra.
However, if we choose instead a different gauge, by setting
$H{z}_{\bar z}=0$, $S=1$,
 the action obtained describes the graviton and
the gauge degree of freedom $\rho{\bar z}_\theta$.
In this gauge, the equation of motion derived from the anomaly
equation is, $$\partial3_{\bar z}H{\bar z}_\theta=0.\eqno(1.8)$$
This equation  is nothing more than the supersymmetrization of the
bosonic gravity equation of motion $\partial_{\bar z}3
h{\bar z}_z$.  It is easily realized that one obtains an $N=1,$
$SL(2,R)$ current  algebra, where the $SL(2,R)$ currents of the
bosonic case  [1] are promoted to superfields with
fermionic components related  to the extra degree of freedom
$\rho{\bar z}_\theta$. This corresponds to the model considered in
[5,6].

 The  $(1,0)$ geometric action, constructed on the coadjoint orbit
of purely central extension of the superVirasoro group is given by
[5,6,11],
$$S_{s.vir}{(1,0)}=-{k\over4\pi}\int
d2zd\theta\Big({\partial_{\bar z}  Z_0+\Theta_0\partial_{\bar
z}\Theta_0\over{(D_{\theta}\Theta_0)}2}\Big)
\Big({D_{\theta}4\Theta_0\over
D_{\theta}\Theta_0}-3{D_{\theta}3\Theta_0\over D_{\theta}\Theta_0}
{D_{\theta}2\Theta_0\over  D_{\theta}\Theta_0}\Big),\eqno(1.9)$$
where $Z_0(z,\bar z,\theta)$ and $\Theta_0(z,\bar z,\theta)$ are
superdiffeomorphisms in $(z,\theta)$ superspace obeying the
superconformal condition,  $$D_{\theta}Z_0=\Theta _0
D_{\theta}\Theta_0,\eqno(1.10)$$
where $D_{\theta}=\partial_\theta+\theta\partial_z$ is the
superderivative.  By
using the superconformal condition, the geometric action (1.9) can
now be rewritten in the form,  $$S_{s.vir}{(1,0)}={k\over2\pi}\int
d2zd\theta{\partial_{\bar z}\Theta_0
D_{\theta}3\Theta_0\over(D_{\theta}\Theta_0)2}.\eqno(1.11)$$

The $(1,1)$ geometric action can be deduced simply by replacing
$\partial_{\bar z}$ with the superderivative
$D_{\bar\theta}=\partial_{\bar\theta}+\bar\theta\partial_{\bar
z}$ in Eq. (1.11),

$$S_{s.vir}{(1,1)}={k\over 2\pi}\int
d2zd2\theta{D_{\bar\theta}\Theta
D_{\theta}3\Theta\over(D_{\theta}\Theta)2}.\eqno(1.12)$$
The relation of the above action to that of $(1,1)$ supergravity
[12] is obtained through the following set of transformations,
$$X(\phi,\bar z,\psi,\bar\theta)=z,\quad \Theta(\phi,\bar
z,\psi,\bar\theta)=\theta.\eqno(1.13)$$  The $(1,1)$ supergravity
action is then given as [12, 25],
$$S(Hz_{\bar\theta})=-{k\over2\pi}\int d2zd2\theta
{\partial_z\psi\over (D_{\theta}\psi)2}\Big(D_{\bar\theta}-
{D_{\bar\theta}\phi-\psi D_{\bar\theta}\psi\over
(D_{\theta}\psi)2}\partial_z\Big)D_{\theta}\psi,\eqno(1.14)$$ where
the gravitational field
is parametrized as, $$Hz_{\bar\theta}={D_{\bar\theta}\phi-\psi
D_{\bar\theta}\psi\over (D_{\theta}\psi)2}.\eqno(1.15)$$

It has been argued in [9], that the geometric actions (1.11) and
(1.12) are equivalent to those obtained by gauging the Borel
subgroup of $(1,0)$ and $(1,1)$ super $SL(2,R)$
Wess-Zumino-Novikov-Witten  models respectively.  However, in the
analysis of [9], it is not quite clear  how the  superconformal
condition can be obtained from the gauged WZNW model.

In [10], the geometric action (1.11) was shown to be
equivalent to the Borel gauged $OSp(1,2)$ WZNW model. There it was
found  that the constraints obtained after integrating out the gauge
field are  equivalent to the superconformal  condition.

Following the analysis of [10], we shall show that the action
(1.12) describes a theory equivalent to a Borel gauged $(0,1)$
$OSp(1,2)$ WZNW model. We will also explain the origin of the
relation (Eq. (1.13)) which connects the geometric action (1.12)
and $(1,1)$ supergravity action and derive the
supergravitational \lq\lq composition formula" [13]. The
relation between the central charge of the $OSp(1,2)$ current
algebra of $(1,1)$ induced supergravity and the conformal anomaly
of the superconformal matter is also obtained.

This work is organised as follows. In section 2, the formulation of
$(1,1)$ suprgravity in the superlight cone gauge [1,7,8] is briefly
reviewed. In section 3 we show the equivalence of the Borel gauged
$(0,1)$ $OSp(1,2)$ WZNW model to that of $S{(1,1)}_{s.vir}$.
Section 4 is concerned in showing the relation of the
geometric action to that of $(1,1)$ supergravity. In the last
section a canonical derivation of the renormalization of the
central charge of the current algebra of supergravity is given.

\vskip 1cm
\centerline{\bf{2. Review of $(1,1)$ Supergravity}}
\vskip 0.2 cm
In this section the theory of $(1,1)$
supergravity formulated in the super light-cone gauge is briefly
reviewed. The $(1,1)$ supergravity theory is described by a set of
covariant derivatives [7, 20]  $$\nabla _A= E_AM D_M +\omega _A
M,\eqno(2.1)$$  where $E_AM$ are the vielbeins,  $w_A$ are the spin
connections and $M$ is the Lorentz generator. The
superspace derivative  $D_A$ is,
$$D_A=\Big(\partial_{z}, \partial_{\bar z}, D_\theta={\partial
\over \partial \theta } +\theta \partial_{ z}, D_{\bar\theta}=
{\partial \over \partial \bar\theta }+\bar\theta \partial_{\bar
z}\Big).$$  The constraints in the $(1,1)$ supergravity
theory are,
 $$ \{ \nabla _\theta, \nabla _\theta \} = 2
\nabla_{ z},\eqno(2.2a)$$ $$ \{ \nabla _{\bar\theta},\nabla
_{\bar\theta} \} =2\nabla _{\bar z},\eqno(2.2b)$$ $$ \{ \nabla
_\theta, \nabla _{\bar\theta}\} =RM. \eqno(2.2c)$$    As a
consequence of the constraints, some of the components $E_AM$
and $\omega_A$ are expressed in terms of a set of independent
superfields. After solving the constraints and going to the super
light-cone gauge, it is found that the only degree of freedom in the
theory is $H_{\bar\theta}{ z}$
and the covariant derivatives are expressed in terms of
$Hz_{\bar\theta}$ as,
$$\eqalign{\nabla_{\bar\theta}&=D_{\bar\theta} -Hz_{\bar
\theta}\partial_z +{1\over2}(D_{\theta} Hz_{\bar \theta})D_\theta-
\partial_z Hz_{\bar \theta}M,\cr
\nabla_\theta &=D_\theta.\cr}\eqno(2.3)$$
The equation of motion for the field $Hz_{\bar\theta}$, derived
from the anomaly equation is,
$$D_{\theta}\partial_z2Hz_{\bar\theta} =0.\eqno(2.4)$$

The solution to the equation of motion (2.4) can be expressed
in terms of superfield currents as,
$$\eqalign{Hz_{\bar\theta}(z,\bar z,\theta, \bar\theta)=&{\cal
J}{+1}(\bar z,\bar\theta)-2{\cal J}0(\bar z,\bar\theta) z+{\cal
J}{-1}(\bar z,\bar\theta)z2 \cr &+\theta\Big({\cal
J}{1\over2}(\bar z,\bar\theta)-{\cal J}{-{1\over2}}(\bar
z,\bar\theta)z\Big).\cr}\eqno(2.5)$$

{}From the calculation of the Ward identity involving the
gravitational field $Hz_{\bar\theta},$ it can be deduced that these
currents satisfy an $N=1$ super $OSp(1,2)$ Kac-Moody algebra [21],
$${\cal J}a({\bar z}_1,{\bar\theta}_1) {\cal
J}b({\bar z}_2,{\bar\theta}_2)= -{{\hat k} /2 \eta {ab} \over
{\bar Z}_{12}} + {{\bar\theta}_{12} f{ab}_c {\cal J}c({\bar
z}_2,{\bar\theta}_2) \over{\bar Z}_{12}},\eqno(2.6)$$ where $\hat k$
is the central charge of the current algebra, ${\bar Z}_{12}=
{\bar z}_1 -{\bar z}_2
-{\bar\theta}_1 {\bar\theta}_2$ and
${\bar\theta}_{12}={\bar\theta}_1-{\bar\theta }_2,$
${{\eta}{ab}}$ is the $OSp(1,2)$ invariant metric and $f{ab}_c$
are the structure constants of $OSp(1,2)$ algebra.
After fixing the super light-cone gauge, the theory is invariant
only under residual transformations respecting the gauge choice. It
can be shown that these residual transformations are generated by
$T{total}$, ${\cal J}{-1}$ and ${\cal J}{-{1\over2}}$, where,
$$T{total}=T{s.matt}+T{s.grav}+T{ghost}\eqno(2.7)$$
is the sum of the contributions of the superconformal matter, the
gravitational field and the ghost fields to the stress energy tensor.
It has to be noted that $T{s.grav}$
has, besides the Sugawara form [24], an additional linear term
in the current ${\cal J}0,$
$$T{s.grav}=T{sug}+\partial_{\bar z} {\cal J}0.\eqno(2.8)$$
This modified stress tensor satisfies a superVirasoro algebra with
conformal anomaly
$$c{s.grav}=\Big({1\over2}+{2k\over
2k+3}\Big)-6(k+{3\over2});\quad \  k=-{\hat
k}-{3\over2}\eqno(2.9)$$
 where the first term comes from the Sugawara form [24] and the
second term derives from the linear term.

 The central charge of the current algebra is determined  by the
coefficient which multiplies the induced quantum supergravity
action and supergravitational quantum fluctuations may lead to a
finite renormalization of this  central charge. To calculate this
requires summing up all loop  contributions to the effective action.
Instead, however, one can use [1,8] the self consistency relation
arising from the  residual invariance of the theory.  This relation
is satisfied if the total central charge  of the theory vanishes,
$i.e$, $$\eqalign{c{total}&=c{s.matt}+c{s.grav}+c{ghost}\cr &=
c{s.matt}+{1\over2}+{2k\over 2k+3}-6(k+{3\over2})-9\cr
&=0\cr}\eqno(2.10)$$
where $-9$ is the sum of the conformal
anomalies of the ghosts  introduced to fix the super light-cone
gauge and $c{s.matt}={3d/2}$ is the conformal anomaly of the
superconformal matter inducing quantum supergravity.

Details of the study of $(1,1)$ supergravity in the
super light-cone gauge are given in [8].
\vskip 2mm
\vfil\eject
\centerline {\bf{3. Borel Gauged $(0,1)$ $OSp(1,2)$ WZNW model}}
\vskip 0.2 cm
We start by reviewing the Borel gauged  $OSp(1,2)$
WZNW model [10]. The resulting model
is equivalent to the geometric action related to $(1,0)$ induced
supergravity.\hfill\break
Let $OSp(1,2)$ denote the set of graded matrices $M$ satisfying,
$$Mt\sigma_1+{(-1)}\kappa\sigma_1 M=0\eqno(3.1)$$ where $\kappa=0$
for the even elements of $M$ and $\kappa=1$ for the odd elements,
$\sigma_1$ is given by,  $$\sigma_1=\pmatrix{0&-1&0\cr 1&0&0\cr
0&0&1\cr}\eqno(3.2)$$ and $Mt$ denotes the supertranspose
of the graded matrix $M$. Eq. (3.1) is solved by,
$$M=\pmatrix{m&n&-\zeta\cr p&-m&\chi\cr
\chi&\zeta&0\cr}\eqno(3.3)$$

where $\{m,n,p\}$ are even and $\{\zeta,\chi\}$ are odd elements.

The  following  basis of $OSp(1,2)$
is now considered,
$$\eqalign{l_{0}&=\pmatrix{1\over2&0&0\cr 0&-{1\over2}&0\cr
0&0&0\cr} l_{1}=\pmatrix{0&1&0\cr 0&0&0\cr
0&0&0\cr}l_{-1}=\pmatrix{0&0&0\cr  1&0&0\cr 0&0&0\cr}\cr
l_{-{1\over2}}&=\pmatrix{0&0&0\cr 0&0&1\cr 1&0&0\cr}
l_{1\over2}=\pmatrix{0&0&1\cr 0&0&0\cr 0&-1&0\cr}.\cr}\eqno(3.4)$$

The Borel subgroup is defined as
the subgroup generated by  $l_{-1}$ and $l_{-{1\over2}}$. The Borel
gauged  $OSp(1,2)$ WZNW  model is described by the following action,
$$S_{gauged}=S(g)+\hbox{str}\int d2z V_1(\partial_z
gg{-1}-K_1),\eqno(3.5)$$ where $g(z,\bar z)\in OSp(1,2)$, str
denotes the supertrace and $S(g)$ is the original $OSp(1,2)$ WZNW
model [16].  The supermatrix $V_1$ is a gauge field taking values in
the Borel algebra of $OSp(1,2)$ and plays the role of a Lagrange
multiplier  and $K_1$ is a constant supermatrix,
$$V_1=\pmatrix{0&0&0\cr A&0&\eta\cr \eta&0&0\cr}\qquad
K_1=\pmatrix{0&1&0\cr 0&0&0\cr 0&0&0\cr}.\eqno(3.6)$$  The
$OSp(1,2)$ group element $g$ can be represented by,
$$g=\pmatrix{1&0&0\cr F_1&1&e_1\cr  e_1&0&1\cr}\pmatrix{F_2&0&0\cr
0&F_2{-1}&0\cr  0&0&1\cr}\pmatrix{1&F_3&e_2\cr 0&1&0\cr
0&-e_2&1\cr}\eqno(3.7)$$ where $\{F_1, F_2, F_3\}$ are bosonic
fields while  $\{e_1, e_2\}$ are fermionic fields.

In expressing the gauged action in terms of the components
fields it is very useful to use the
Polyakov-Wiegmann identity  [15],
$$S(g_1g_2)=S(g_1)+S(g_2)+{k\over2\pi}\int d2z
\hbox{str}\bigg(g_1{-1}\partial_{\bar z} g_1\partial_z g_2
g_2{-1}\bigg).\eqno(3.8)$$
 Integrating over the gauge field in (3.5)
imposes the constraints,
$$F_22(\partial_z F_3+\partial_z e_2
e_2)=1,\eqno(3.9a)$$
$$F_2 \partial_z e_2-e_1=0,\eqno(3.9b)$$
 which, after being substituted back into the gauged action and using
the Borel local gauge invariance to gauge away the field $\Phi$,
leaves a reduced theory which can be expressed only in terms of the
dynamical fields $F_3$ and $e_2$ in the following reduced form,
  $$S(F_3, e_2)={k\over
2\pi}\int d2z\bigg(F_2{-
1}\partial_z F_2F_2{-1}\partial_{\bar z} F_2- F_22\partial_z
e_2\partial_{\bar z}\partial_z e_2 \bigg)\eqno(3.10)$$  supplemented
by (3.9a).

We now turn to the action (1.11) and express it in
component form. In terms of the components of $\Theta_0$,
$$\Theta_0=\alpha_0(z,\bar z)+\theta\beta_0(z,\bar z),\eqno(3.11)$$
the action (1.11) can be expressed as,
$$S_{s.vir}{(1,0)}={k\over2\pi}\int d2z  \Big({\partial_{\bar
z}\beta_0\partial_z\beta_0\over\beta_02}+{\partial_z\partial_{\bar z}
\alpha_0\partial_z\alpha_0\over\beta_02} \Big).\eqno(3.12)$$
If we now make the identification
$$\beta_0={1\over F_2},\quad \alpha_0=ie_2,
\eqno(3.13)$$
then the action (3.10) coincides with (3.12).

The action (1.11) is supplemented with the superconformal
condition (1.10)
 which, in terms of the components of $\Theta_0$ and $Z_0=X_0+\theta
X_1,$ gives,

$$\partial
_z X_0=-\alpha_0\partial_z\alpha_0+\beta_02.\eqno(3.14)$$
Making  the identification  $X_0=F_3$, (3.14) becomes the
constraint (3.9a) which supplements the reduced action (3.10).

Now it is straightforward to generalize the above analysis to
the Borel gauged $(0,1)$ $OSp(1,2)$  WZNW model. Here the model is
described by three (0,1) bosonic
superfields $\{{\cal F}_1, {\cal F}_2, {\cal F}_3\}$ and two
fermionic (0,1) superfields $\{{\cal E}_1, {\cal
E}_2\}$.

The Borel gauged $(0,1)$ $OSp(1,2)$ WZNW model is described by the
following action $$S_{gauged}=S(G)+\hbox{str}\int d2zd\bar\theta
V(\partial_z GG{-1}-K_1).\eqno(3.15)$$
where $G(z,\bar z,\bar\theta)$ is a group element of $OSp(1,2)$,
$S(G)$ is the original  $(0,1)$ $OSp(1,2)$ WZNW model and
the supermatrix $V_1$ is a gauge superfield taking values in
the Borel algebra of $OSp(1,2)$ and plays the role of a Lagrange
multiplier.

In expressing the gauged action in terms of the components fields,
it is very useful to employ the supersymmetric version of
the Polyakov-Wiegmann identity,
$$S(G_1G_2)=S(G_1)+S(G_2)+{k\over2\pi}\int d2z d\bar\theta
\hbox{str}\bigg(G_1{-1}D_{\bar\theta}G_1\partial_z G_2
G_2{-1}\bigg).\eqno(3.16)$$ Following the same steps as for the
bosonic case, we express the gauged action in terms of the
dynamical superfields ${\cal F}_3$ and ${\cal E}_2$ in the following
 reduced form,
$$S({\cal F}_3, {\cal E}_2)={k\over 2\pi}\int
d2zd\bar\theta\Big({\cal F}_2{-1}\partial_z{\cal F}_2{\cal
F}_2{-1}D_{\bar\theta}{\cal F}_2-
{\cal F}_22\partial_z{\cal E}_2\partial_z D_{\bar\theta}{\cal E}_2
\Big),\eqno(3.17)$$  supplemented by,  $${\cal
F}_22(\partial_z{\cal F}_3+\partial_z {\cal E}_2 {\cal E}_2)
=1.\eqno(3.18)$$ The gauged action is invariant under the right
symmetry $G\rightarrow GQ'(\bar z,\bar\theta).$  We represent
$Q'(\bar z,\bar\theta)\in OSp(1,2)$ by, $$Q'(\bar
z,\bar\theta)=\pmatrix{a&b&\alpha\cr c&d&\beta\cr
a\beta-c\alpha&b\beta- d\alpha&1+\beta\alpha\cr}\eqno(3.19)$$
where $\{a,b,c,d\}$ are even and  $\{\alpha, \beta\}$ are odd
elements, with $(ad- bc=1-\beta\alpha).$  Therefore, the reduced
action  is invariant under, $$\eqalign{{\cal F}_2&\rightarrow {d
{\cal F}_2+b\over c {\cal F}_2+a}+ {(\beta {\cal F}_2+\alpha){\cal
E}_2\over{(c {\cal F}_2+a)}2},\cr {\cal E}_2&\rightarrow
{\alpha+\beta
{\cal F}_2\over c{\cal F}_2+a}+{
{\cal E}_2\over  c{\cal F}_2+a}.\cr}\eqno(3.20)$$
These are the super $OSp(1,2)$ transformations.
The maximal symmetry in the left moving part of the theory is
the superconformal symmetry [19].

In conclusion, the Borel gauged
$(0,1)$ $OSp(1,2)$ model has the same symmetries as
the $S{(1,1)}_{s.vir}$ model, $i.e.$, a super $OSp(1,2)$ Kac-Moody
symmetry  in the right-moving sector and superdiffeomorphism in the
the left moving part of the theory.

 We now turn to the action (1.12) and express it in terms
of the component fields. Expanding $\Theta=\Theta_0(z,\bar
z,\theta)+ \bar\theta\Theta_1(z,\bar z,\theta),$
the action (1.12) can be expressed in the form,
$$S_{s.vir}{(1,1)}={k\over 2\pi}\int d2zd\theta
\Big({\partial_{\bar z}\Theta_0\partial_z
D_{\theta}\Theta_0\over{(D_{\theta}\Theta_0)}2}+{\partial_z\Theta_1
D_{\theta}\Theta_1\over {(D_{\theta}\Theta_0)}2}
\Big).\eqno(3.21)$$ In terms of the components of  $\Theta_0$ and
$\Theta_1,$

$$\Theta_0=\alpha_0(z,\bar z)+\theta\beta_0(z,\bar z),\eqno(3.22a)$$
$$\Theta_1=\alpha_1(z,\bar z)+\theta\beta_1(z,\bar z),\eqno(3.22b)$$
the action (3.21) becomes,
$$S_{s.vir}{(1,1)}={k\over 2\pi}\int d2z
\Big({\partial_{\bar
z}\beta_0\partial_z\beta_0\over\beta_02}+{\partial_z\partial_{\bar
z} \alpha_0\partial_z\alpha_0\over\beta_02}-
2{\partial_z\alpha_0\partial_z\alpha_1\beta_1\over\beta_03}+
({\partial_z\alpha_1\over\beta_0})2+
{\partial_z\beta_1\beta_1\over\beta_02}\Big).\eqno(3.23)$$

We also express  the reduced action (3.17) in components.
Writing
 ${\cal F}_2=\lambda_1+\bar\theta\lambda_2$ and
${\cal E}_2=f_1+\bar\theta  f_2,$ the reduced action (3.17) is
expressed,
$$\eqalign {S({\cal F}_3,{\cal E}_2)={k\over 2\pi}\int
d2z\Big(&{\partial_z\lambda_1\partial_{\bar
z}\lambda_1+\partial_z\lambda_2\lambda_2\over\lambda_12}-2{\lambda_1\lambda_2
\partial_z f_1\partial_z f_2}\cr &+{\lambda_12\partial_z
f_1\partial_{\bar z}\partial_z f_1}-\lambda_12(\partial_z
f_2)2\Big).\cr}\eqno(3.24)$$

Now making the identification,
$$\beta_0={1\over\lambda_1},\quad \alpha_0=if_1,\quad
\alpha_1=if_2, \quad
\beta_1={\lambda_2\over\lambda_12},\eqno(3.25)$$ the action
(3.24) is found to coincide with (3.23).

The action (1.12) is supplemented with the superconformal
condition,
$$D_{\theta}Z=\Theta D_{\theta}\Theta;\quad Z=Z_0+\bar\theta Z_1$$
which gives, $$D_{\theta}Z_0=\Theta_0
D_{\theta}\Theta_0,\eqno(3.26a)$$ $$D_{\theta}Z_1=-\Theta_0
D_{\theta}\Theta_1-\Theta_1D_{\theta}\Theta_0.\eqno(3.26b)$$
Expanding $Z_0$ and $Z_1$ as follows,
$$Z_0=X_0+\theta X_1, \qquad Z_1=Y_0+\theta Y_1,$$ Eq. (3.26)
yields,
$$\partial X_0=-\alpha_0\partial\alpha_0+\beta_02,\eqno(3.27a)$$
$$\partial Y_0=-2\beta_0\beta_1-
\alpha_1\partial\alpha_0+\alpha_0\partial\alpha_1.\eqno(3.27b)$$

The  constraint (3.18) supplementing the reduced action (3.17),
in components, gives,
$$\lambda_12(\partial_z F_1+\partial_z f_1f_1)-1=0,\eqno(3.28a)$$
$$\lambda_12(\partial_z F_2+\partial_z f_2f_1-\partial_z
f_1f_2)+2{\lambda_2\over\lambda_1}=0,\eqno(3.28b)$$
where ${\cal F}_3=F_1+\bar\theta F_2$. Identifying $F_1$
and $F_2$ with $X_0$ and $Y_0$ respectively and using  (3.25),
equations (3.27a) and (3.27b) are shown to coincide with (3.28a) and
(3.28b) respectively.  The above demonstration shows the equivalence
of the two  theories. \vskip 0.5 cm

\centerline{\bf{4. Gauge transformations and superdiffeomorphisms}}
\vskip 0.2 cm
In the introduction it was stated that the $(1,1)$ supergravity
action (Eq. (1.14)) is connected to  the action $S_{s.vir}{(1,1)}$
through a change of variables (Eq. (1.13)).  In this
section, this connection is explained by exploiting the relation
between superdiffeomorphisms and restricted gauge transformations
[13]. We first derive the $(1,1)$ supersymmetric version of the
Polyakov-Wiegmann identity, the so called \lq\lq composition
formula" [15]. Then from the relation between superdiffeomorphisms
and restricted gauge transformations [13], the connection
between  the geometric action $S_{s.vir}{(1,1)}$ and
$S_{s.grav}{(1,1)}$ is verified, appearing to arise from a
consequence  of a simple property of WZNW models. Finally  the
\lq\lq composition formula" for the $(1,1)$ supergravity action is
obtained, which is  also  reminiscent of that for the WZNW model.

   Consider a (1,1) supersymmetric two dimensional gauge theory
with an $OSp(1,2)$ gauge group. This theory is described in
terms of two  gauge fields $A_\thetaa$ and $A_{\bar\theta}a$, where
$a=\Big({1,0,-1,{1\over2}, -{1\over2}}\Big)$ is an $OSp(1,2)$ gauge
group index. The gauge transformations of the gauge fields are given
as,
$$\delta A_\thetaa=-{\cal D}_\theta\epsilona=
-\Big(D_{\theta}\epsilona-fa_{bc}A_\thetab\epsilonc\Big),\eqno(4.1a)$$
$$\delta A_{\bar\theta}a=-{\cal D}_{\bar\theta}\epsilona
=-\Big(D_{\bar\theta}\epsilona-
fa_{bc}A_{\bar\theta}b\epsilonc\Big),\eqno(4.1b)$$ where
$fa_{bc}$ are the structure constants of the $OSp(1,2)$ algebra.

The action of the gauge theory described by
$A_\theta$, is given by,
$$S_1(A_\theta)\sim \hbox{log}\ \hbox{sdet}
(D_{\theta}-A_\theta).\eqno(4.2)$$  Its variation under gauge
transformations is now,
$$\delta S_1(A_\theta)=\int
d2zd2\theta\ \hbox{str}(J_{\bar\theta}\delta A_\theta),\eqno(4.3)$$
where $J_{\bar\theta}$ is the gauge current satisfying the anomaly
equation  $${\cal D}_\theta
J_{\bar\theta}=-{k}D_{\bar\theta}A_\theta.\eqno(4.4)$$  By
substituting (4.1a) in (4.3) and using the above anomaly equation, we
obtain

$$\eqalign{\delta S_1(A_\theta)&=-\int d2zd2\theta
\ \hbox{str}\Big(J_{\bar\theta}{\cal D}_{\theta}\epsilon\Big)\cr
&=-\int d2zd2\theta \ \hbox{str}\Big(\epsilon{\cal
D}_{\theta}J_{\bar\theta}\Big)\cr &=k\int d2zd2\theta
\ \hbox{str}(\epsilon{D_{\bar\theta}}A_\theta).\cr}\eqno(4.5)$$
We now Parametrize $A_\theta$ by,$$ A_\theta =D_{\theta}g
g{-1},\eqno(4.6)$$ where $g$ is a group  element of $OSp(1,2)$
gauge group in (1,1) superspace. Then $S_1(A_\theta)$ is solved by a
(1,1) super WZNW model, $S_1(g)$.

Similarly one can parametrize
$A_{\bar\theta}=D_{\bar\theta}hh{-1}$, where $h$ is a group element
of $OSp(1,2)$ gauge group in (1,1) superspace, and study the
dynamics of the gauge field $ A_{\bar\theta},$ to find that,
$$S_2(A_{\bar\theta})\sim \hbox{log}\ \hbox{sdet}
(D_{\bar\theta}-A_{\bar\theta})\eqno(4.7)$$ is also solved by an
$N=1$ $OSp(1,2)$ WZNW model, $S_2(h).$  The final form of the
total effective action is then,
$$S_{eff}(g,h)=S_1(g)+S_2(h)-k\int
d2zd2\theta \hbox{str}(D_\theta
gg{-1}D_{\bar\theta}hh{-1}),\eqno(4.8)$$ where the last term is
added to insure gauge invariance. Now, a finite gauge transformation
on $A_{\theta}$ and $A_{\bar\theta}$ is given by,
$$g\rightarrow Ug,\qquad h\rightarrow Uh, \quad U\in
OSp(1,2).\eqno(4.9)$$ As the  effective action (4.8) is invariant
under this transformation, this implies the following relation,

$$S_{eff}(g,h)=S_{eff}(Ug,Uh).\eqno(4.10)$$
If we set $U=h{-1}$ or $U=g{-1}$, we can then deduce
that
$$S_{eff}(g,h)=S_1(h{-1}g)=S_2(g{-1}h),\eqno(4.11)$$
and in particular,
$$S_1(h{-1})=S_2(h)\eqno(4.12).$$
Finally, we arrive at the (1,1) supersymmetric extension of the
Polyakov-Weigmann composition formula,
$$S_1(h{-1}g)=S_1(g)+S_1(h{-1})-k\int d2zd2\theta
\ \hbox{str}(D_\theta gg{-1}D_{\bar\theta}hh{-1}).\eqno(4.13)$$

Following Polyakov [13], we now partially fix a gauge
by imposing the following conditions,
$$A_\theta1=A_\theta0=A_\theta{-{1\over2}}=0, \quad
A_{\theta}{1\over2}=1.\eqno(4.14)$$ Then the gauge transformations
(4.1), give the following relations,
$$\delta
A_\theta{-1}=-D_{\theta}\epsilon{-1}
+A_\theta{-1}\epsilon0,\eqno(4.15a)$$
$$\delta A_\theta{1}=-D_{\theta}\epsilon{1}-
2\epsilon{1\over2}=0,\eqno(4.15b)$$
$$\delta
A_\theta{0}=-D_{\theta}\epsilon{0}-2A_\theta{-1}\epsilon1+2
\epsilon{-{1\over2}}=0,\eqno(4.15c)$$
$$\delta
A_\theta{1\over2}=-D_{\theta}\epsilon{1\over2}-{1\over2}\epsilon0
=0,\eqno(4.15d)$$
$$\delta A_\theta{-{1\over2}}=-D_{\theta}\epsilon{-{1\over2}}+
A_\theta{-1}\epsilon{1\over2}-\epsilon{-1}=0,\eqno(4.15e)$$ which
then give the following relations among the gauge parameters,
$$\epsilon{0}=-2D_{\theta}\epsilon{1\over2}=\partial_z\epsilon{1},\eqno(4.16a
)$$
$$\epsilon{-1}={1\over2}A_\theta{-1}D_{\theta}\epsilon{1}-
D_{\theta}A_\theta{-1}
\epsilon{1}-{1\over2}\partial_z\epsilon0.\eqno(4.16b)$$

These relations, when substituted back into (4.15a) now give,
$$\delta A_\theta{-1}=+{1\over2}D_{\theta}\partial2_z\epsilon1+
{1\over2}D_{\theta}A{-1}_\theta
D_{\theta}\epsilon1 +\partial_z
A_\theta{-1}\epsilon1+{3\over2}A_\theta{-1}\partial_z\epsilon1.
\eqno(4.17)$$
If we set $T_{\theta z}=kA_\theta{-1},$ we then obtain
the following equation
$$\delta
T_{\theta
z}={k\over2}D_{\theta}\partial2_z\epsilon1+
{1\over2}D_{\theta}T_{\theta
z}D_{\theta}\epsilon1 +\partial_z T_{\theta z}\epsilon1+
{3\over2}T_{\theta z}\partial\epsilon1.\eqno(4.18)$$
This relation represents the infinitesimal superconformal
transformation of the superstress energy tensor. This is because
within the background of $A_\theta{1\over2}=1$, the isospin is
equivalent to the spin and thus all the fields in this
background will acquire an additional spin equals to  their
$OSp(1,2)$ isospin [13]. Therefore $A{-1}_\theta$ is a spin $3/2$
superfield and also, $\epsilon1$ can be identified with
$\epsilonz.$   In the partially gauge fixed theory discussed above,
the WZNW model will become a model describing the dynamics of
$T_{\theta z}$ and so $\delta S(A_\theta)$
becomes,
$$\delta
S(T_{\theta z})=
\int d2zd2\theta \epsilonz D_{\bar\theta} T_{\theta
z},\eqno(4.19)$$ where $\delta S(T_{\theta z})$ represents the
transformation under the superconformal transformation (4.18).
In addition, $\delta S(T_{\theta z})$ is given by,
  $$\delta
S(T_{\theta z})=-\int d2zd2\theta Lz_{\bar\theta}\delta T_{\theta
z},\eqno(4.20)$$ where $Lz_{\bar\theta}$ is a function of
$T_{z\theta}$. Comparing the above two equations and using (4.18),
we deduce that $Lz_{\bar\theta}$ should satisfy the following
equation,  $$\Big(D_{\bar\theta}-Lz_{\bar\theta}\partial_z
+{1\over2}(D_\theta Lz_{\bar\theta})D_\theta-
{3\over2}(\partial_z  Lz_{\bar\theta})\Big)T_{\theta z}
=-{k\over2}D_\theta\partial2_z Lz_{\bar\theta}.\eqno(4.21)$$

Define $S(Hz_{\bar\theta})$ as the Legendre transform of
$S(T_{\theta z})$ [13], then its transformation under
superdiffeomorphisms is given by,
$$\delta S(Hz_{\bar\theta})=\int
d2zd2\theta Z_{\theta z}\delta Hz_{\bar\theta},\eqno(4.22)$$
where $Z_{\theta z}$ satisfies the following equation,
$$\Big(
D_{\bar\theta}-Hz_{\bar\theta}\partial_z+
{1\over2}(D_{\theta}Hz_{\bar\theta})D_\theta-
{3\over2}(\partial_z Hz_{\bar\theta})\Big)Z_{\theta z}
=-{k\over2}D_{\theta}\partial2_zHz_{\bar\theta},\eqno(4.23)$$
and the field $Hz_{\bar\theta}$ satisfies the following
transformation,
$$\delta Hz_{\bar\theta}=
\Big(
D_{\bar\theta}-Hz_{\bar\theta}\partial_z+{1\over2}(D_{\theta}
Hz_{\bar\theta})D_{\theta}-(\partial_z
Hz_{\bar\theta})\Big)\epsilonz.\eqno(4.24)$$
Finally, we define the action
$$W(Hz_{\bar\theta}, T_{\theta z})=S (Hz_{\bar\theta})+S(T_{\theta
z})-\int d2zd2\theta Hz_{\bar\theta}T_{\theta z}.\eqno(4.25)$$
It can be easily checked that this combined action is invariant
under the transformations given by equations (4.18) and (4.24).

We turn now to find a solution for the action
$S(Hz_{\bar\theta})$. Parametrize $Hz_{\bar\theta}$ as,
$$Hz_{\bar\theta}={ D_{\bar\theta}\phi-\psi D_{\bar\theta}\psi\over
(D_\theta\psi)2},$$ where $\phi$ and $\psi$
satisfy,$$D_\theta\phi=\psi D_\theta\psi,\eqno(4.26)$$ then the
anomaly equation (4.21) is solved by a superSchwartzian derivative,
$$Z_{\theta z}=k\Big({\partial2_z\psi\over D_\theta\psi}-2
{\partial_z\psi\partial_zD_\theta\psi\over
(D_\theta\psi)2}\Big)=k{\cal S}(\psi),\eqno(4.27)$$ and the action
$S(Hz_{\bar\theta})$ is given by,
$$S(Hz_{\bar\theta})=-kS_{s.grav}{(1,1)}(\phi,\psi)=-k\int
d2zd2\theta {\partial_z\psi\over (D_\theta\psi)2}\Big(
D_{\bar\theta}-
 {D_{\bar\theta}\phi-\psi D_{\bar\theta}\psi\over
(D_\theta\psi)2}\partial_z\Big)D_\theta\psi.\eqno(4.28)$$

The action $S(T_{\theta z})$ is the geometric quantization of
the superVirasoro algebra, and it is given by,

$$\eqalign{S(T_{\theta z})&=kS_{s.vir}{(1,1)}=k\int
d2zd2\theta\Big({D_{\bar\theta}\Theta
D_\theta3\Theta\over(D_\theta\Theta)2}\Big)\cr &=-{k\over2}\int
d2zd2\theta\Big({ D_{\bar\theta}X-\Theta D_{\bar\theta}\Theta\over
(D_{\theta}\Theta)2}\Big) \Big({\partial2_z\Theta\over
D_\theta\Theta}-3 {\partial_z\Theta\partial_zD_\theta\Theta\over
(D_\theta\Theta)2}\Big),\cr}\eqno(4.29)$$ where $X(z,\bar
z,\theta,\bar\theta)$ and $\Theta(z,\bar z,\theta,\bar\theta)$
satisfy the superconformal condition
$$D_\theta X=\Theta D_\theta\Theta, \eqno(4.30)$$ and
$$\eqalign{Lz_\theta &=
{D_{\bar\theta}X-\Theta D_{\bar\theta}\Theta\over
(D_\theta\Theta)2},\cr
T_{\theta z}&=k\Big({\partial2_z\Theta\over D_\theta\Theta}-
2{\partial_z\Theta\partial_zD_\theta\Theta\over
(D_\theta\Theta)2}\Big)=k{\cal S}(\Theta).\cr}\eqno(4.31)$$

The finite form of (4.18) and (4.24) can be represented by
$$\eqalign{X(z,\bar z,\theta,\bar\theta)&\rightarrow X(X_1,\bar
z,\Theta_1,\bar\theta),\cr
\Theta(z,\bar z,\theta,\bar\theta)&\rightarrow \Theta(X_1,\bar
z,\Theta_1,\bar\theta),\cr
\phi(z,\bar z,\theta,\bar\theta)&\rightarrow \phi(X_1,\bar
z,\Theta_1,\bar\theta),\cr
\psi(z,\bar z,\theta,\bar\theta)&\rightarrow \psi(X_1,\bar
z,\Theta_1,\bar\theta),\cr}\eqno(4.32)$$
which symbolically will be written as
$$(X, \Theta)\rightarrow (X, \Theta)\bullet(X_1,\Theta_1),\quad
(\phi,\psi)\rightarrow
(\phi,\psi)\bullet(X_1,\Theta_1).\eqno(4.33)$$
These relations are obviously the gravitational analogue of,
$$g\rightarrow Ug,\qquad h\rightarrow Uh.$$
Also $(X{-1},\Theta{-1})$ is defined as,
$$(X,\Theta)\bullet(X{-1},\Theta{-1})=(z,\theta).\eqno(4.34)$$
This to be understood as the gravitational analogue of
$gg{-1}=I$, where $I$ is the identity.

The invariance of the combined action
$W(Hz_{\bar\theta}, T_{\theta z})$ under superdiffeomorphisms
implies the relationship,
$$W\Big((X, \Theta),(\phi,\psi)\Big)=
W\Big((X,
\Theta)\bullet(X_1,\Theta_1),(\phi,\psi)\bullet(X_1,\Theta_1)\Big).
\eqno(4.35)$$
If we set
$(X_1,\Theta_1)=(\phi{-1},\psi{-1})$ or
$(X_1,\Theta_1)=(X{-1},\Theta{-1})$, then the above equation gives
$$\eqalign{W\Big((X,\Theta),(\phi,\psi)\Big)&=
kS_{s.vir}{(1,1)}\Big((X,\Theta)\bullet(\phi{-1},\psi{-1})\Big)
\cr
&=-kS_{s.grav}{(1,1)}\Big((\phi,\psi)\bullet(X{-1},\Theta{-1})\Big),\cr}\eqno
(4.36)$$
and in particular, $$kS_{s.vir}{(1,1)}\Big((X,\Theta)\Big)
=-kS_{s.grav}{(1,1)}\Big((X{-1},\Theta{-1})\Big).\eqno(4.37)$$
This last relation is  reminiscent of
$$S_1(h{-1})=S_2(h),$$
and explains why one obtains the classical supergravity action from
the geometric action when $$(X,\Theta)\bullet(\phi,\psi)=(z,
\theta).\eqno(4.38)$$
 Finally we obtain the \lq\lq composition formula",
$$\eqalign{kS_{s.vir}{(1,1)}\Big((X,\Theta)
\bullet(\phi{-1},\psi{-1})\Big)&=-kS_{s.grav}{(1,1)}
\Big((\phi,\psi)\Big)
+kS_{s.vir}{(1,1)}\Big((X,\Theta)\Big)\cr &- k\int
d2zd2\theta{D_{\bar\theta}\phi-\psi D_{\bar\theta}\psi\over
(D_\theta\psi)2}{\cal S}(\Theta).\cr}\eqno(4.39)$$

\vskip 0.6 cm

\centerline{\bf{5. Induced Quantum (1,1) Supergravity}}
\vskip 0.2 cm
In this section we study the quantum theory of $(1,1)$ supergravity
using the analysis derived in the previous sections. Initially, it
has been demonstrated that the classical geometric action describing
the superVirasoro algebra is equivalent to the constrained (0,1)
$OSp(1,2)$ WZNW model. However, it has been argued by Ooguri and
Bershadsky [19] that the Borel gauged bosonic $OSp(1,2)$ WZNW model
at level $k$ gives a theory whose  left moving part gives
representations of the superVirasoro algebra with  central charge,
$$c_k={3d_k\over2}={15\over2}-3(2k+3)-{3\over(2k+3)},\eqno(5.1)$$
while the right moving part is described by  an
$OSp(1,2)$ current algebra.
It can  easily be deduced  that the Borel gauged $(0,1)$
$OSp(1,2)$ WZNW at level $k$ has an $N=1$ $OSp(1,2)$ current
algebra in the the right moving sector while the
left moving sector is described by a superconformal field
theory with central charge given by Eq.(5.1)

Therefore, it can be concluded that quantum mechaniclly, the theory
of the superVirasoro group with the geometric action   $-(d_k/4)
S{(1,1)}_{s.vir}(\Theta)$ is equivalent to the constrained
 $(0,1)$ $OSp(1,2)$ WZNW model of level $k$. The modification
of the central charge which multiplies $S{(1,1)}_{s.vir}(\Theta)$
(which is $k$ at the classical level) could also be understood in
terms of the Jacobian factor resulting from the change to a
superdiffeomorphic invariant measure in the functional integral of
$S{(1,1)}_{s.vir}(\Theta)$.

The superstress energy tensor generating the
superVirasoro algebra satisfies the operator product expansion
$$T(Z_1)T(Z_2)={d_k/4\over
Z_{12}3}+3/2{\theta_{12}T(Z_2)\over
Z_{12}2}+{\theta_{12}\partial_{{z_2}} T(Z_2)\over Z_{12}} +1/2
{D_{{\theta_2}}T(Z_2)\over Z_{12}},\eqno(5.2)$$ where $Z$ denotes a
point in $(1,1)$ superspace, $Z_{12}= z_{1}
-z_{2}-\theta_{1}\theta_{2}$ and $\theta_{12}=\theta_{1}-\theta_{2}$.

 In the super light-cone gauge, the effective $(1,1)$ supergravity
action is the generating functional for the superstress energy
tensor of the supermatter inducing it. The operator product
expansion of the superstress tensor implies the following relation
for its generating functional $\Gamma(Hz_{\bar\theta})$,  $$\Big(
D_{\bar\theta}-Hz_{\bar\theta}\partial_z+{1\over2}(D_{\theta}
Hz_{\bar\theta})D_{\theta}-{3\over2}(\partial_z
Hz_{\bar\theta})\Big){\delta\over\delta
Hz_{\bar\theta}}\Gamma(Hz_{\bar\theta})=-
{d_k\over8}\partial_z2 D_{\theta} Hz_{\bar\theta}.\eqno(5.3)$$
This is solved by,  $$\Gamma(Hz_{\bar\theta})={d_k\over4}
S{(1,1)}_{s.vir}(\Theta),\eqno(5.4)$$ where, $$Hz_{\bar
\theta}={D_{\bar\theta} \phi-\psi D_{\bar\theta} \psi\over
({D_{\theta}\psi}2)},\quad \Theta(\phi,\bar
z,\psi,\bar\theta)=\theta.\eqno(5.5)$$

Therefore, the quantum (1,1) supergravity is defined by the
following functional integral,
$$\int [dHz_{\bar \theta}]\exp
\Big({id_k\over4}S_{s.vir}{(1,1)}(\Theta)\Big).\eqno(5.6)$$ We will make a
change of variable in the above functional integral as follows,
$$Hz_{\bar \theta}+\delta Hz_{\bar\theta}
={D_{\bar\theta} \phi(z+\delta z,\bar z,\theta+\delta
\theta,\bar\theta)- \psi(z+\delta z,\bar
z,\theta+\delta\theta,\bar\theta)D_{\bar\theta}
 \psi(z+\delta z,\bar
z,\theta+\delta\theta,\bar\theta)\over ({D_{\theta}
\psi(z+\delta z,\bar
z,\theta+\delta\theta,\bar\theta)})2}.\eqno(5.7)$$ By using the
relations
$$\eqalign{\phi(z+\delta z,\bar z,\theta+\delta\theta,\bar\theta)&=
\phi(z,\bar z,\theta,\bar\theta)+\epsilonz\partial_z\phi+
{1\over2}D_{\theta}\epsilonzD_{\theta}\phi,\cr
\psi(z+\delta z,\bar z,\theta+\delta\theta,\bar\theta)&=
\psi(z,\bar z,\theta,\bar\theta)+\epsilonz\partial_z\psi+
{1\over2}D_{\theta}\epsilonzD_{\theta}\psi,\cr}\eqno(5.8)$$
where$$\epsilonz=\delta z+\theta\delta\theta$$ is an infinitesimal
superdiffeomorphism parameter,
we can then write
$$Hz_{\bar \theta}+\delta
Hz_{\bar\theta}=Hz_{\bar \theta}+\Big(D_{\bar\theta} -Hz_{\bar
\theta}\partial_z  +
{1\over2}(D_{\theta} Hz_{\bar \theta})D_{\theta}-
\partial_z
Hz_{\bar \theta}\Big)\epsilonz.\eqno(5.9)$$
Therefore, we now have,
$$\eqalign{[dHz_{\bar\theta}]&=\hbox{sdet}
\Big(D_{\bar\theta} -Hz_{\bar\theta}\partial_z
+{1\over2}(D_{\theta}Hz_{\bar \theta})D_{\theta}
-\partial_z Hz_{\bar
\theta}\Big)[d\epsilonz]\cr  [dHz_{\bar\theta}]&= \exp
\Big(-i{10\over4}S_{s.vir}{(1,1)}(\Theta)\Big)[d\epsilonz].\cr}
\eqno(5.10)$$
Thus, the functional integral of (1,1) supergravity can be written
 as
$$\int [d\epsilonz]\exp
\Big(-i{10-d_{k}
\over4}S_{s.vir}{(1,1)}(\Theta)\Big).\eqno(5.11)$$
By noticing that $10-d_k=d_{-(k+3)}$, we deduce that
induced quantum supergravity theory is equivalent to a Borel
gauged $(0,1)$ $OSp(1,2)$ WZNW model at level $k'=-(k+3)$. The
relation between the level of the current algebra, $ k'$ and $d_k$
agrees with the result  calculated
 using the residual invariance of the theory (Eq. 2.10).

In conclusion, we have demonstrated the equivalence of
$S{(1,1)}_{s.vir},$ the geometric action describing superconformal
field theory to the theory obtained by gauging the Borel subgroup
of the $(0,1)$ supersymmetric $OSp(1,2)$ WZNW model. The action
$S{(1,1)}_{s.vir}$ is also understood as a partially gauged $(1,1)$
supersymmetric $OSp(1,2)$ WZNW model in which the residual gauge
group transformation becomes the superconformal transformation of the
superstress energy tensor of $S{(1,1)}_{s.vir}.$ The classical
action of $(1,1)$ supergravity in the super light-cone gauge
is then derived as a Legendre transform of $S{(1,1)}_{s.vir}$
and the relation between the two actions is a consequence of the
properties of WZNW models.

The relation of $(1,1)$ supergravity to that of WZNW model is
then exploited to derive the renormalization of the central charge of
the current algebra of $(1,1)$ supergravity.

The WZNW models with symmetry group $\cal G$ can be constructed as
two dimensional field theories defined in terms of the standard
Kirillov-Kostant  symplectic two form [22] on the coadjoint orbit
of the $\cal G$ Kac-Moody group. Therefore, the relation between
$S{(1,1)}_{s.vir}$ and the super $OSp(1,2)$ WZNW model
reflects the fact that the symplectic structure of the
coadjoint orbit of the superVirasoro  group is related  to that of
the $OSp(1,2)$ coadjoint orbit via Hamiltonian reduction [2,4].

The above analysis can be extended to the more interesting case of
$N=2$ supergravity, in which case the supergravity is related to the
$N=2$ supersymmetric $OSp(2,2)$ WZNW model. The currents of the $N=2$
super Kac-Moody algebra satisfy non-linear chirality constraints
[17] which could be related  to the non-linearity of the covariant
derivatives in the super light-cone gauge formulation of $(2,2)$
supergravity [18].  We shall report on that in a seperate
publication.
\vskip 1cm

\centerline{ ACKNOWLEDGEMENT}

I would like to thank  M.B. Green, C. Hull, T. Kuramoto,
 B. Spence for encouragement and useful
conversations. \vskip 1.5 cm
\centerline {REFERENCES}
\vskip 4mm
1. A.M. Polyakov, Mod. Phys. Lett. A2 (1987) 893; \hfill\break
V.G. Knizhnik, A.M. Polyakov and A.B. Zamolodchikov, Mod. Phys.
Lett. \hfill\break
A3 (1988) 819; \hfill\break
A.M. Polyakov and A.B. Zamolodchikov, Mod.  Phys. Lett. A3
(1988)1213;\hfill\break
Lectures given by Polyakov at Les Houches summer school on Fields,
Strings  \hfill\break
and Critical Phenomena (1988).\hfill\break
2. A. Alekseev and Shatashvilli, Nucl. Phys B323 (1989)
719.\hfill\break
3. E. Witten, Comm. Math. Phys. 114 (1988)
1.\hfill\break  4. M. Bershadsky and H. Ooguri Comm. Math. Phys.
2(1989)49.\hfill\break   5. S. Aoyama, Phys. Lett. 228B (1989) 335,
S. Aoyama and  J. Julve, preprint\hfill\break
DFPD/89/TH/60.
\hfill\break
6. H. Aratyn, E. Nassimov, S. Pacheva and S. Solomon,
Phys. Lett. 234B (1990) 307.\hfill\break
7. M.T. Grisaru and R.M. Xu, Phys.
Lett. 205B (1988) 486.\hfill\break
8. T. Kuramoto, Nucl. Phys B346 (1990) 527.\hfill\break
9. S. Aoyama, preprint Padova DEPD/89/TH/75.\hfill\break
10. W.A. Sabra, Mod. Phys. Lett. A6 (1991) 875.\hfill\break
11. G. Delius, P. van Nieuwenhuizen and V.G.J. Rodgers,
Int. J. Mod. Phys. A5 (1990) 3943.\hfill\break
12. J. Grundberg and R. Nakayama, Mod. Phys. Lett. A4 (1989)
55.\hfill\break
13. A. M. Polyakov, Int. J. Mod. Phys. A5 (1990) 833.\hfill\break
14. V.G. Kac and T. Todorov, Comm. Math. Phys. 102 (1985)
337.\hfill\break
15. A.M. Polyakov and P.B. Wiegmann, Phys. Lett. 141B
(1984) 233.\hfill\break
16. E. Witten, Comm. Math. Phys. 92 (1984) 455.\hfill\break
17. C.M. Hull and B. Spence, Phys. Lett. B241 (1990).\hfill\break
18. A. Alnowaiser, Class. Quantum Grav. 7 (1990) 1033.\hfill\break
19. M. Bershadsky and H. Ooguri, Mod. Phys. Lett, 229B
(1989) 374.\hfill\break
20. S.J. Gates, M.T. Grisaru, L. Mezincescu and P.K. Towensend,
Nucl. Phys. B286 (1987) 1.\hfill\break
21.  V.G. Kac and T. Todorov, Comm. Math. Phys. 102 (1985) 337; Y.
Kazama and H. Suzuki, Nucl. Phys. B321 (1989) 232.\hfill\break
22. A.A.  Kirillov, Elements of the Theory of  Representations,
Springer Verlag (1976). \hfill\break
23. R. Brooks, F. Muhammad and
S.J. Gates, Nucl. Phys. B268  (1986) 599.\hfill\break
24. J. Fuchs, Nucl. Phys. B286  (1987) 455.\hfill\break
25. F. Delduc and F. Gieres, preprint CERN-TH 5953/90.\hfill\break
\bye